\documentclass[prl,twocolumn,showpacs,preprintnumbers,amsmath,amssymb]{revtex4}
\usepackage{graphicx}
\begin{document}
\title{Hydrogenated grain boundaries in graphene     }
\author{W. H. Brito}
\author{R. Kagimura}
\author{R. H. Miwa}
\affiliation{Instituto de F\'{\i}sica, Universidade Federal de
Uberl\^andia, CP 593, 38400-902, Uberl\^andia, MG, Brazil.}
%\email{}

\date{\today}

\begin{abstract}
  We have investigated by means of first principles calculations the
  structural and electronic properties of hydrogenated graphene
  structures with distinct grain boundary defects.  Our total energy
  results reveal that the adsorption of a single H is more stable at
  grain boundary defect.  The electronic structure of the grains
  boundaries upon hydrogen adsorption have been examined.  Further
  total energy calculations indicate that the adsorption of two H on
  two neighbor carbons, forming a basic unit of graphane, is more
  stable at the defect region.  Therefore, we expect  that these
  extended defects would  work as a nucleation region for the formation
  of a narrow graphane strip embedded in graphene region.

\end{abstract}

\pacs {73.22.-f, 73.22.Pr, 71.55.-i }

%73.22.-f 	Electronic structure of nanoscale materials and related systems
%73.22.Pr 	Electronic structure of graphene 
%71.55.-i 	Impurity and defect levels
\maketitle

Physical and chemical properties of hydrogenated two-dimensional
graphene-based (nano)structures have been the aim of recent
researches. The adsorption of H atoms can modify the structural,
electronic and magnetic properties of the ideal graphene, which is a
null-gap semiconductor~\cite{graphene}. It is known that the
adsorption of a single H at graphene opens an energy gap and
introduces spin-polarized gap states~\cite{hallmark}. On the other
hand, a fully hydrogenated graphene~\cite{graphane1} is an insulator
material~\cite{graphane2}, known as graphane.  Partially hydrogenated
graphene structures, such as graphene/graphane
superlattices~\cite{superlattice}, graphene nanoroads~\cite{road} and
graphane nanoribbons~\cite{graribbon} have been theoretically
proposed, and can be produced by removing H atoms from
graphane~\cite{sessi} or adsorbing them on
graphene~\cite{haberer,balog}.  These structures can present a
suitable band gap energy~\cite{haberer,balog} for pratical
applicantions in the new nanoelectronic devices.

Grain boundary (GB) defects have been observed in graphene in very
recent experiments~\cite{gb57nature,gbnature}.  Theoretical
works~\cite{gblouie} have also proposed structural models for GB
defects, which resemble to those of Refs.~\cite{gb57nature,gbsimonis}.
Tight-binding calculations suggest that the H adsorption is more
stable at the GB defects than in perfect graphene~\cite{gbtrend}.
However, there is no \emph{ab-initio} investigations on the properties
of H adsorption at those observed GB defects, H/GB.  Thus, in this
work we have investigated, by means of first-principles calculations,
the effect of H adsorption on the properties of graphene with grain
boundary defects.  Our results indicate that the adsorption of a
single H is more stable at GB defect.  Also, the adsorption of two H
on two neighbor carbons, forming a basic unit of graphane, is more
stable at the defect region.  Thus, we can infer that these extended
defects would work as a nucleation region for the formation of a
narrow graphane strip embedded in graphene region.  The modifications
in the electronic structure due to H adsorption at GB defects are also
investigated.

Spin-polarized density functional theory (DFT)~\cite{ks}  calculations, 
within the generalized-gradient approximation (GGA)~\cite{gga}, 
were performed using the SIESTA  code~\cite{siesta}, where core 
states were replaced by norm-conserving  pseudopotentials~\cite{pse-tm} 
in the factorized form~\cite{KL}.  We have employed a double-zeta plus
 polarization (DZP)  basis set with an energy shift of 100 meV~\cite{base}, and 
an energy  cutoff of 200 Ry for the real-space mesh. 
10  (210) special k-points for geometry (band structure) calculations 
%for  the Brillouin Zone  integration in 
%the Monkhorst-Pack~\cite{MP} scheme 
were used.
All atomic positions were fully relaxed until the 
residual forces were converged to within 10~meV/\AA.  We have employed
a   distance of  15~\AA~ between graphene  layers
and  supercells  with (60) 60, 96, and 50 carbons   atoms to represent
the (perfect)  defective graphene structures.
% shown in the figures: ~\ref{grainmodel}~(a),
%~\ref{grainmodel}~(b), and ~\ref{grainmodel}~(c), respectively.

\begin{figure}[t]
\includegraphics[height=6cm]{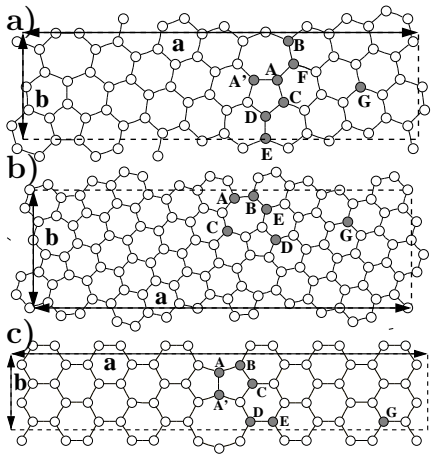}
\caption{Optimized geometries  of distinct grain
boundary defects, labeled as: (a) GB1(5-7), (b) GB2(5-7), and (c) GB(5-8). 
The dashed-line  box indicate their periodic unit cells.
The  unit cell parameters a and b, in \AA,  are 24.5 and 6.6; 28.9 and 9.0; and 27.3 and 5.0
for  GB1(5-7), GB2(5-7), and  GB(5-8), respectively.  The  adsorption sites are in gray.}
\label{grainmodel}
\end{figure}

We first describe the atomic structure of GB defects
in graphene. Here, we have considered three distinct structural models
for the GB defects, shown in Fig.~\ref{grainmodel}. These extended
defects can be built from a suitable regular arrangement of 5-, 7- or
8-membered carbon rings along a line in graphene.  The GB structure
shown in  Fig.~\ref{grainmodel}~(a), named as GB1(5-7), is formed
by an alignment of pentagon-heptagon pair separated by
hexagons~\cite{gbsimonis}. Figure.~\ref{grainmodel}~(b) shows the GB
structure named as GB2(5-7), which is a regular sequence of
pentagon-heptagon pair~\cite{gblouie}.  In these structures, due to
the periodicity of the supercell, we have considered two GBs in the unit
cell. The third GB structure, shown in the Fig.~\ref{grainmodel}~(c),
is named as GB(5-8) and consists of a sequence of pentagon-pairs and
octagon along the graphene zig-zag direction~\cite{gbnature}.

Let us now investigate the energetic stability of H atoms adsorbed at
graphene with GB defects, H/GB.  We first consider a single H atom adsorbed
on the top of several non-equivalent carbon atoms at the defect
region. These adsorption sites are indicated by the letters A, B, C,
D, E and F, as shown in  Fig.~\ref{grainmodel}.  Also, we have
considered the H adsorption at the graphene-like region (site G). To
investigate that, we calculated the adsorption energy, $E_{ads}$,
which is defined by the following equation:
\begin{equation}
E_{ads}=E[{\rm{H/GB}}]-E[{\rm{GB}}]-E[{\rm{H}}]+\delta_{{\rm{BSSE}}}
\end{equation}
where $E[{\rm{H/GB}}]$ ($E[{\rm{GB}}]$) is the total energy of the
fully relaxed hydrogenated (pristine) graphene structure with GB
defects, and $E[\rm H]$ is the total energy of the atomic hydrogen.
The last term ($\delta_{{\rm{BSSE}}}$) is used to correct the basis
set superposition error~\cite{bsse}. A negative value for $E_{ads}$
indicates that the adsorption of H is an energetically favourable
process.  To test our methodology, we have calculated the adsorption
energy for a single H at perfect graphene. The calculated adsorption
energy of -0.81 eV/H atom is in excelent agreement with other
\emph{ab-initio} calculations~\cite{spillover,hgraphite, prl93}.

\begin{table}[!h]
\centering
\caption{Calculated adsorption energies, in eV/H atom, for a single H atom adsorbed at graphene with
grain boundary defects. The letters indicate the adsorption sites (see Fig.~\ref{grainmodel}). }
\vspace{0.30cm}
\label{eads}
\begin{tabular}{lccc}
\hline \hline
    \:&   H/GB1(5-7)  \:&  H/GB2(5-7)    \:& H/GB(5-8)   \\
\hline
 G  \:&   -0.83       \:&   -0.93     \:&    -0.78   \\
 A  \:&   -1.55       \:&   -1.75     \:&    -2.29   \\
 B  \:&   -1.39       \:&   -1.54     \:&    -1.89   \\
 C  \:&   -1.30       \:&   -1.45    \:&     -1.59   \\
 D  \:&   -1.26      \:&    -1.24     \:&    -1.43   \\
 E  \:&   -1.18      \:&    -0.99     \:&    -1.31   \\
 F  \:&   -0.96      \:&              \:&           \\
\hline
\hline
%---------------------------------------------------------------------%
\end{tabular}
\end{table}

Our calculated adsorption energies, shown in the Table~\ref{eads},
indicate that $E_{ads}$ at the graphene-like region (site G) are
within 0.1 eV to that at the perfect graphene sheet.  Thus, the error
due to the size cell effect is small compared to our calculated energy
differences.  It is interesting to note that the adsorption energy at
the GB defect is negative and smaller than that at perfect graphene
for all investigated adsorption sites. This indicates that the
adsorption of a single H is energetically more stable and likely to
occur at the defective region.  Indeed, we find that  H/GB(5-8) is
a quite stable structure, since the hydrogen adsorption energy
(2.29~eV/H) is comparable with the binding energy of an isolated H$_2$
molecule (2.16~eV/H).
%{\bf The lowest adsorption energy
%  (site A), relative to that at perfect graphene, ranges from -0.7 to
%  -1.5 eV/H, which is at the same order to that of a single H adsorbed
%  at the Stone-Wales defect in graphene, -1.2 eV/H atom~\cite{swh}.  O
%  que se conclui daqui? Ou qual a conec\c{c}\~ao com o resto do
%  texto?}  
The flatness of pristine graphene structures is lost due to the H
adsorption.  For H/GB2(5-7), the hydrogenated carbon and its nearest
neighbors move upwards by 0.5 and 0.2~\AA, respectively, while the
other atoms exhibit negligible displacements.  Meanwhile, for the
H/GB1(5-7) and H/GB(5-8) systems, the hydrogenated carbon moves
upwards by 0.4~\AA~ relative to its nearest neighbors, and a
significant displacement is viewed even for the fourth nearest atom of
the hydrogenated carbon.  At the equilibrium geometry, the C--H bond
lenght is 1.13~\AA\ and the bond length of hydrogenated carbon and its
nearest neighbors ranges from 1.49 to 1.55 \AA, indicating a sp$^3$
hybridization for the carbon atom bonded to H.

\begin{figure}[t]
\includegraphics[height=9cm]{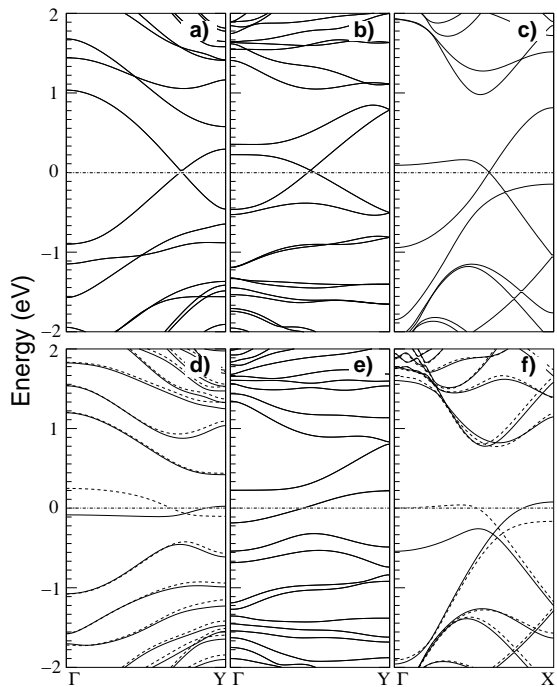}
\caption{
Calculated band structure for pristine and hydrogenated graphene with GB defects.
a) and d)  GB1(5-7), b) and e),    (c) and f) GB(5-8). Solid (dashed) lines indicate the spin-up (spin-down) components.}
\label{banda}
\end{figure}

\begin{figure}[t]
\includegraphics[height=10cm]{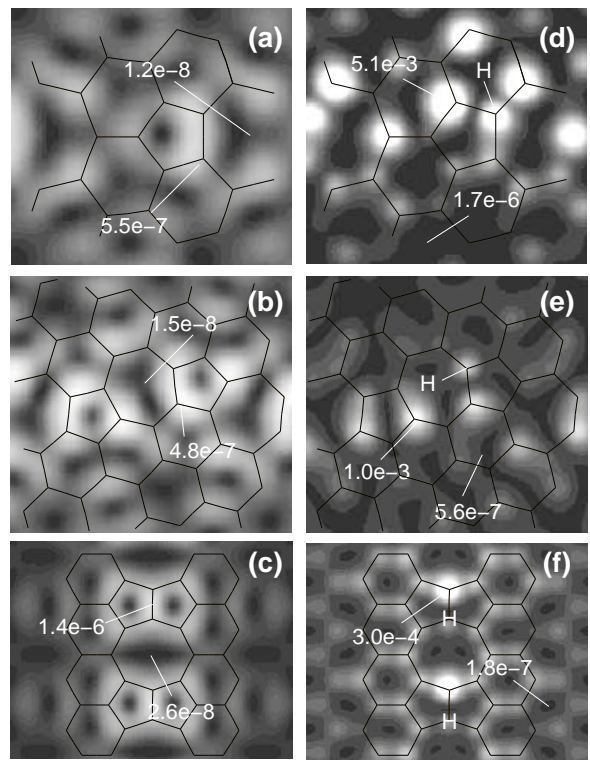}
\caption{(Color online) Simulated STM images for the pristine (left panels) and
for hydrogenated (right panels) GB defects (whitin an energy interval of ($E_{f} - 0.2$) eV and $E_{f}$).
The densities are in $electrons/bohr^{3}$.}
\label{stm}
\end{figure}

Focusing on the electronic properties of the H/GB systems, we
performed band structure calculations and scannning tunneling
microscopy (STM) simulations for the most stable configurations.
Figures~\ref{banda}~(a)--(c) present the electronic band structure of
the pristine GB1(5-7), GB2(5-7), and GB(5-8) systems,
respectively. The valence and conduction bands exhibit a linear
energy--momentum relation nearly of the Dirac point for wave vectors
parallel to the GBs.  Our STM simulations, for the occupied states
within an energy window of 0.2~eV below the calculated Fermi level
[Figs.~\ref{stm}~(a)--(c)], indicate the formation of bright lines
parallel to the GBs, which is in accordance with the more reactive
character of the GBs in comparison with the perfect graphene
sheet. Those findings are in good agreement with previous
works~\cite{gbsimonis,gblouie,gbnature}.  Figures~\ref{banda}~(d)--(f)
show the electronic band structure of H/GB1(5-7), H/GB2(5-7), and
H/GB(5-8) structures, respectively. Our results show that the single H
adsorption has a strong effect on the electronic properties of GB
defects. The systems remain metallic, however, the Dirac point
disappears, and it is noticeable the formation of spin-polarized
states crossing the Fermi level for the hydrogen adsorbed GB1(5-7) and
GB(5-8) systems.  In contrast, although the structural similarity with
the H/GB1(5-7) system, H/GB2(5-7) exhibits a spin paired state crossing
the Fermi level.

The STM images of the GBs are modified upon hydrogen
adsorption. Figures~\ref{stm}~(d)--(f) show alternating bright and
dark spots along a direction parallel to the GBs. The presence of H
adatoms somewhat localizes the electronic states, which is in
accordance with the small energy dispersion of the electronic states
(near the Fermi level) as depicted in Figs.~\ref{banda}~(d)--(f). In
H/GB1(5-7) we find the formation of bright spots on the hydrogen
adsorbed site as well as on the nearby C atoms, Fig.~\ref{stm}~(d).
Meanwhile, the STM image of the H adatom becomes faded for H/GB2(5-7)
and H/GB(5-8). In H/GB(5-8), Fig.~\ref{stm}~(f), we find a very weak
spot lying on the H adatom, while the nearest neighbor C atom
(parallel to the GB) becomes brighter. Indeed, our calculated
projected density of states for H/GB(5-8) indicate that the electronic
states along the C--H bond are mostly localized between 2 and 3~eV
below the Fermi level, while in H/GB1(5-7), the electronic states of H
adatoms (mostly) lie at $E_F \pm 0.3$~eV.  In this case, in contrast
with the H/GB1(5-7) system, the contribution of the H adatom in
H/GB(5-8) is not relevant for the spin polarized states near the Fermi
level, Fig.~\ref{banda}~(f).  The STM results for the H/GB2(5-7)
system indicate that the electronic states are less localized,
Figs.~\ref{stm}~(e).  The hydrogen adsorbed sites as well as the
nearest neighbor C atoms present almost the same brightness along the
GB.

% A similar result is found for the spin paired state of GB2(5-7). In
% contrast, for GB(5-8), the main contribution for the spin polarized
% states is due to the C atom bounded to the hydrogenated carbon.
%  In contrast,
% for the H/GB1(5-7) system, the most of the electronic states along the
% C--H bonds lie at 0.5~eV (????) below the Fermi level. The STM results
% for the H/GB2(5-7) system indicate that the electronic states, within
% $E_F - 0.2$~eV, are less localized, Figs.~\ref{stm}(e)--(h).  In this
% case, the hydrogen adsorbed sites as well as the nearest neighbor C
% atoms present almost the same brightness along the GB.

% The STM images of the hydrongen adsorbed GBs, Figs.~\ref{stm}(c), (f),
% and (i),  show alternating bright and dark spots along a direction
% parallel to the GBs. Therefore, the main contribution for those spin
% polarized states of GB1(5-7) is due to the bright carbon atoms and the
% H atom. A similar result is found for the spin paired state of
% GB2(5-7). In contrast, for GB(5-8), the main contribution for the spin
% polarized states is due to the C atom bounded to the hydrogenated
% carbon.  In this case, the contribution of H is not relevant.  To
% confirm the STM results, we have calculated the density of states
% projected onto the brighest C and H atoms shown in the
% Fig. ~\ref{stm}.  Our results confirm that the contribution of those C
% and H atoms to the total density of states is significant near the
% valence band maximum.

Finally, we consider the adsorption of two H atoms on graphene
structures. We focus on the most stable configurations for the single
H adsorption.  Thus, we place the H atoms on sites A and A' for
GB1(5-7) and GB(5-8), and sites A and B for GB2(5-7).  The
configuration with the H atoms sitting on the opposite sides of the
sheet is energetically more favorable than that on the same side.
Also, our calculated adsorption energy on the perfect graphene, -1.7
eV/H atom, is in excellent agreement with other \emph{ab-initio}
calculations~\cite{spillover}.  For the defective graphene region, we
find adsorption energies of -1.86, -1.98 and -2.12~eV/H atom for
GB1(5-7), GB2(5-7) and GB(5-8), respectively. Therefore, the
adsorption of two H atoms is more stable at the defect region.  The
structural configuration of the hydrogenated carbons resembles the
basic unit of graphane.  The above results allow us to infer that the
formation of graphane stripes, embedded in graphene, can be ruled by
the presence of grain boundary defects.

In conclusion, we have investigated the effect of adsorption of H
atoms in the electronic and structural properties of graphene with
(extended) grain boundary defects.  Our {\it ab-initio} calculations
indicate that the GB defect could work as nucleation region for
graphane strip formation.  The electronic structure of such
hydrogenated defect is strongly affect by the adsorption of a single H
atom. Spin polarized states are observed in H/GB1(5-7) and H/GB(5-8)

% latex nome
% bibitex nome 
% latex nome 
% latex nome
%\bibliography{RHMiwa}

%\end{document}

\begin{acknowledgments}
We acknowledge support from the Brazilian agencies CNPq and FAPEMIG.
\end{acknowledgments}

\end{document}